\documentstyle[twocolumn,aps,psfig]{revtex}

\begin{document}
\draft
 
\twocolumn[\hsize\textwidth\columnwidth\hsize\csname
@twocolumnfalse\endcsname

\title{Exact Quantization of Even-Denominator 
Fractional Quantum Hall State
at $\nu=5/2$ Landau Level Filling Factor}

\author{W. Pan$^{1,2}$, J.-S. Xia$^{2,3}$, 
V. Shvarts$^{2,3}$, E.D. Adams$^{2,3}$, H.L. Stormer$^{4,5}$,
D.C. Tsui$^{1}$, L.N. Pfeiffer$^5$, K.W. Baldwin$^5$, 
and K.W. West$^5$}

\address{
$^1$Princeton University, New Jersey 08544}
\address{
$^2$NHMFL, Tallahassee, Florida 32310}
\address{
$^3$University of Florida, Gainesville, Florida 32611}
\address{
$^4$Columbia University, New York, New York 10027}
\address{
$5$Bell Laboratories, Lucent Technologies, Murray Hill, New Jersey 07974}

\date{\today}
\maketitle

\begin{abstract} 
We report ultra-low temperature experiments on the obscure fractional 
quantum Hall effect (FQHE) at Landau level 
filling factor $\nu=5/2$ in a very high mobility specimen 
of $\mu=1.7 \times 10^7$ cm$^2$/Vs. We achieve an electron temperature 
as low as $\sim$ 4~mK, where we observe vanishing $R_{xx}$ and, 
for the first time, a quantized Hall resistance, 
$R_{xy}=h/(5/2e^2)$ to within 2 ppm. $R_{xy}$ at the neighboring 
odd-denominator states $\nu=7/3$ and 8/3 is also quantized. 
The temperature dependences of the $R_{xx}$-minima at 
these fractional fillings yield activation 
energy gaps $\Delta_{5/2}=0.11$~K, $\Delta_{7/3}=0.10$~K, 
and $\Delta_{8/3}=0.055$~K. 
\end{abstract}

\vskip2pc]

Electrons in two-dimensional  systems at low 
temperatures and in the presence of an intense 
magnetic field condense into a sequence of incompressible 
quantum fluids with finite energy gaps for quasiparticle excitation, 
termed collectively the fractional quantum 
Hall effect (FQHE) \cite{qhebooks}. These highly correlated 
electronic states occur at rational fractional 
filling $\nu=p/q$ of Landau levels. Their characteristic 
features in electronic transport experiments are 
vanishing resistance, $R_{xx}$, and exact quantization of 
the concomitant Hall resistance, $R_{xy}$, to $h/(p/qe^2)$. 
Over the years, a multitude of FQHE states 
have been discovered -- all $q$'s being odd numbers. 
The only known exceptions are the states at 
half-filling of the second Landau level $\nu$ = 5/2 (=2+1/2) 
and $\nu$ = 7/2 (=3+1/2) 
\cite{willett:prl87,eisenstein:prl88,gammel:prb88}. 
Half-filled states in the 
lowest Landau level show no FQHE, whereas half-filled 
states in still higher Landau levels exhibit yet 
unresolved anisotropies \cite{anisotropy}. Recent experiments in 
tilted magnetic field even seem to hint at a 
connection between the $\nu=9/2$ state and the state at $\nu=5/2$
\cite{anisotropytilt}.

The origin of the $\nu$ = 5/2 and 7/2 states remains 
mysterious. Observation of odd-denominator 
FQHE states is intimately connected to the 
anti-symmetry requirement for the electronic 
wave function. An early, so-called hollow-core 
model \cite{haldane:prl88} for the FQHE 
at $\nu$ = 5/2 and 7/2, which takes 
explicitly into account aspects of the modified 
single-particle wave functions of the second 
Landau level, arrived at a trial wave function. 
However, for a Coulomb Hamiltonian its 
applicability is problematic \cite{macdonald:prb89,morf:prl98}. 

With the advent of the composite fermion (CF) 
model \cite{cfbooks} the existence of exclusively 
odd-denominator FQHE states is traced back 
to the formation of Landau levels of CFs 
emanating from even-denominator fillings, 
such as the sequence $\nu=p/(2p \pm 1)$ from $\nu=1/2$. 
Even-denominator fillings themselves represent 
Fermi-liquid like states, resulting from 
the attachment of an even number of magnetic 
flux quanta to each electron. The obvious conflict 
between this theory and experiment at $\nu=5/2$ is 
resolved by invoking a CF-pairing mechanism 
\cite{greiter:prl91,moore:npb91,park:prb98,nick:prl99}. 
In loose analogy to the formation of Cooper 
pairs in superconductivity such pairing creates 
a gapped, BCS-like ground state at $\nu=5/2$, called 
a ``Pfaffian'' state, which displays a FQHE. 
Indeed, an exact numerical diagonalization 
calculation by Morf \cite{morf:prl98} favors the Pfaffian state.

The experimental situation remains poor. 
Study of the $\nu=5/2$ state requires ultra-high 
mobility specimens and its small energy gap 
necessitates ultra-low temperatures. 
Previous experiments 
\cite{willett:prl87,eisenstein:prl88,gammel:prb88} indicated the 
existence of a local minimum in $R_{xx}$ and a 
slope-change in $R_{xy}$ at $\nu=5/2$. However, we still 
lack the observation of the unmitigated, 
well-developed hallmarks of a FQHE at $\nu=5/2$, 
namely vanishing of the magneto-resistance, 
$R_{xx}$, and quantization of the Hall resistance 
to $R_{xy}=h/(5/2e^2)$. Furthermore, the absence of 
vanishing $R_{xx}$ so far prohibited the determination 
of a true activation energy at $\nu=5/2$ and neighboring 
fractions. Instead, one had to rely on 
ad-hoc approximations, such as employing 
the ratio of $R_{xx}$ at the minimum to $R_{xx}$ at 
adjacent peaks, to extract a measure for the 
size of the energy gap.

In this letter, we present ultra-low temperature 
data on the even-denominator FQHE state $\nu = 5/2$ 
and its vicinity. For the first time, we observe 
a wide Hall plateau, precisely quantized to 
$R_{xy}= 2h/5e^2$ to an accuracy better than $2 \times 10^{-6}$. 
Concomitantly, the longitudinal resistance 
assumes vanishing values $R_{xx} = 1.7 \pm 1.7~\Omega$ at 
a bath temperature, $T_b=8$~mK. True activation 
energy measurements, performed between 8~mK and 
50~mK, yield an energy gap of $\Delta_{5/2}=0.11$~K. 
The energy gaps of neighboring FQHE states 
at $\nu=7/3$ and $\nu=8/3$ are $\Delta_{7/3}=0.10$~K 
and $\Delta_{8/3}=0.055$~K respectively.

We used a standard 4mm $\times$ 4mm geometry for 
our samples with eight indium contacts 
diffused symmetrically around the perimeter. 
An electron density of $2.3 \times 10^{11}$~cm$^{-2}$ was 
established by illuminating the sample with light 
from a red light-emitting diode at 4.2~K. The 
sample mobility is $\mu=1.7 \times 10^7$~cm$^2$/Vs 
at $\sim$ 1~K and below. 
Cooling the electron system to ultra-low 
temperatures is a formidable task, since the 
electron-phonon coupling between the 2DES and its 
host lattice decreases precipitously with decreasing 
temperature \cite{gammel:prb88,wennberg:prb86}. An earlier experiment 
on $\nu = 5/2$ \cite{gammel:prb88} reached 
an electron temperature, $T_e$, 
of only 9~mK in spite of a bath temperature of $T_b \sim 0.5$~mK. 
In fact, at very low temperatures cooling of the 
electrons proceeds largely via the electrical contacts. 
Electrons diffuse to the contacts, where they 
cool in the highly disordered region formed by 
the ``dirty'' alloy of GaAs and Indium. Therefore, 
cooling of the contacts is of paramount importance 
in low-temperature transport experiments and our 
cooling system was designed to cool specifically 
the contact areas of the 2DES specimen. Eight sintered 
silver heat exchangers each having an estimated 
surface area of $\sim$ 0.5~m$^2$  and formed around a 10~mil 
silver wire were soldered directly to the indium 
contacts of the sample using indium as a solder. 
They provide electrical contact and simultaneously 
function as large area cooling surfaces. The 
backside of the sample was glued with gallium to 
yet another large surface area heat exchanger for 
efficient cooling of the lattice of the specimen. 
The inset of Fig.~1 shows a sketch of this 
arrangement. Sample and heat exchangers were 
immersed into a cell made from polycarbonate, 
equipped with electrical feed-throughs and 
filled with liquid $^3$He. The liquid is cooled by 
the $PrNi_5$ nuclear demagnetization stage 
of a dilution refrigerator via well-annealed 
silver rods that enter the $^3$He-cell. 
This brings the system's base temperature 
to 0.5~mK, well below the 8~mK of the dilution unit.

A CMN thermometer, a $^3$He melting curve thermometer, 
and/or a Pt-NMR thermometer, are mounted in the 
low-field region, at the top of the nuclear stage. 
The accuracy of the thermometry is estimated to 
be better than 0.05~mK. All measurements were 
performed in an ultra-quiet environment, 
shielded from electro-magnetic noise. RC 
filters with cut-off frequencies of 10~kHz 
were employed to reduce RF heating. The data 
were collected using a PAR-124A analog lock-in 
with an excitation current of 1~nA at typically 5~Hz. 
At this current level electron heating was 
undetectable for temperature greater than 8~mK. 
This was deduced from a series of heating experiments, 
in which Rxx was measured at different currents from 
0.5~nA to 100~nA at $T_b=8.0$~mK. The resistance of the 
strongly $T$-dependent $R_{xx}$ peak at 3.75~T in Fig.1 
was used as an internal thermometer for the 
electron temperature, $T_e$, assuming $T_e=T_b$ at 
higher $T_b$ and in the limit of low current. 
The $T_e$ vs $I$ data fit the expected 
$T^5$-relationship \cite{gammel:prb88,wennberg:prb86}, 
ln($I^2$)= const + ln($T_{e}^{5} - T_{b}^{5}$), where $I$ is in nA, 
$T$ in mK and const = -7.5.  At $T_b = 8.0$~mK and 1~nA 
current it yields $T_e=8.05$~mK, sufficiently close to 
$T_b$ for the difference to be negligible. 

Fig.~1 shows the Hall resistance $R_{xy}$ and the 
longitudinal resistance $R_{xx}$ between Landau level 
filling factors $\nu=3$ and $\nu=2$ at $T_b=2$~mK. 
Using the above equation for an excitation current of 
1nA, the deduced electron temperature, $T_e$, is $\sim$ 4~mK. 
Ultimately it is unclear, whether the $T^5$-law 
continues to hold. However, if anything, 
we would expect $T_e \sim 4$~mK to be a conservative estimate, 
since electron cooling via the well heat-sunk 
contacts should lower $T_e$ below the limit set by 
lattice cooling. Strong minima emerge in 
$R_{xx}$ in the vicinity of filling factors $\nu=5/2$, $\nu=7/3$ 
and $\nu=8/3$. Satellite features can be made out, 
that are probably associated with filling 
factors $\nu$ = 19/7, 13/5, 12/5 and 16/7. If this 
identification is correct, apart from the $\nu=5/2$ state, 
the successive development of FQHE states is 
not unlike the one observed in the lowest 
Landau level. However, we need to caution, 
that as long as such minima do not show 
concomitant Hall plateaus, one cannot be certain 
as to their quantum numbers. Strikingly different 
from the $R_{xx}$ pattern around $\nu=1/2$ is not only the 
existence of the central $\nu=5/2$ state, but 
also the emergence of very strong maxima flanking this minimum. 

The primary minima at $\nu=5/2$, $\nu=7/3$, and $\nu=8/3$ show 
well developed Hall plateaus. In particular the 
5/2-plateau is extensive, allowing its value to 
be measured to high precision. This was performed 
with a current of 20~nA, which raises $T_e$ to $\sim$ 15~mK, 
but increases the signal to noise, while keeping 
the plateau largely intact. The lock-in operated 
at 23~Hz and its output was averaged for 20 min. 
The neighboring integral quantum Hall effect (IQHE) 
plateaus at $\nu=2$ and $\nu=3$ were used as standard resistors 
to which the value at $\nu=5/2$ was compared. $R_{xy}$ was 
found to be quantized to $h/(5/2)e^2$ to better 
than $2 \times 10^{-6}$. It unambiguously establishes the 
state at $\nu=5/2$ as a true FQHE state. The plateau 
values around filling factor $\nu=7/3$ and $\nu=8/3$ were 
derived in a similar fashion. They are quantized to 
their respective $R_{xy}$ values to better than $3.5 \times 10^{-4}$. 

An unusual set of features of the Hall trace are 
pronounced maxima and minima between plateaus. 
Ultimately we do not know their origin. The simplest 
interpretation is a mixing of $R_{xx}$ into $R_{xy}$. 
Indeed, two of the peaks in $R_{xy}$ ($B \sim 3.8$~T and $B \sim 4.0$~T) 
become dips on field reversal, although of smaller 
amplitude than the peaks. The minimum in $R_{xy}$ at $B \sim 3.5$~T 
does not invert on field reversal but remains a dip of 
similar strengths. In any case, these pronounced features 
in $R_{xy}$ do not line up with the strong maxima in $R_{xx}$, 
but seem to be shifted further away from the 
central $\nu=5/2$ state. It is also remarkable, 
that $R_{xy}$ of these maxima and minima approaches 
the Hall resistance of the IQHE at $\nu=2$ and $\nu=3$, 
respectively. The situation is somewhat 
reminiscent of data taken in the regime of the 
anisotropic phase around $\nu=9/2$ and 11/2 \cite{anisotropy}, but 
it is premature to invoke similar mechanisms. 

Having established the existence of correctly 
quantized FQHE states at $\nu =5/2$, as well as 
at $\nu=7/3$ and 8/3, we turn to the measurement 
of their energy gaps. Fig.2 gives an overview 
over the temperature dependence of $R_{xx}$ between 
$\nu=2$ and $\nu=3$ at four different temperatures. 
These data were recorded during a separate cool 
down than the data of Fig.1. Comparison of both 
provides some measure for their reproducibility. 
The predominant fractions at $\nu$ = 7/3, 5/2 and 8/3 
reproduce very well, whereas the secondary features, 
assigned in Fig.~1 to $\nu$ = 19/7, 13/5, 12/5, and 16/7 are 
considerably weaker than there. Accordingly, 
only the gaps of the primary fractions are accessible. 
However, the approach of vanishing $R_{xx}$ in these 
three cases, for the first time enables us to 
perform true activation energy measurements on these states. 

Fig.~3 shows on a semilog graph the value of 
$R_{xx}$ at the position of the $\nu$ = 7/3, 5/2 and 8/3 
minima as a function of inverse temperature 
for 8~mK $<$ ($T_e=T_b$) $<$ 50~mK. The characteristic $S$-shape of 
such data is observed \cite{boebinger:prl85}. The high-temperature 
(low $1/T$) roll-off is a result of the temperature 
approaching the energy gap. The low-temperature 
(high $1/T$) saturation is also a standard feature of 
such data. Its origin has never unambiguously 
been identified, but is most likely associated with 
a transition to hopping conduction. The $\nu=5/2$ data 
show an anomalous feature at low temperature. The 
kink at $T \sim 15$~mK has been observed in all three 
activation measurements, that we performed during 
a half year span, while the sample had been kept 
below 4.2~K. We do not know its origin. In any case, 
for all three fractions there exists a central 
region of data points, that follow an exponential 
dependence for almost one order of magnitude. 
We use a straight line approximation to this 
region to determine the activation energy of 
each fraction and deduce an energy 
gap $\Delta_{\nu}$ from 
$R_{xx}(T) = const \times$ exp($-\Delta_{\nu}/2kT$). 
Their values are included in Fig.~3. 

The energy gap $\Delta_{5/2}=0.11$~K is very close 
to the value deduced earlier from the temperature 
dependence of the depth of the 5/2 minimum 
with respect to the height of the two adjacent 
peaks \cite{eisenstein:prl88}. Since these peaks have themselves a 
strong temperature dependence it must be 
concluded that the rise in peak height is 
comparable to the drop of the 5/2 minimum. 
It relates the two features in a yet unexplained 
fashion. The energy gaps of the two $p/3$ states 
are $\Delta_{7/3}=0.10$~K and $\Delta_{8/3}=0.055$~K. This 
implies $\Delta_{7/3} \approx \Delta_{5/2}$ 
and $\Delta_{8/3} \approx \Delta_{5/2}/2$. 
At first glance such a relationship between 
gaps in the second Landau level is very 
satisfying, since it closely reproduces the 
relationship between the calculated energy 
gaps \cite{morf:prl98}, assuming the 5/2 state to be of 
the Pfaffian type. However, with 
$\Delta_{5/2}^{theo} \approx 0.02~e^2/\varepsilon \lambda_0 \approx 1.9$~K
at 3.67~T, the absolute values in theory and in 
experiment differ by more than an order of 
magnitude. This is a common observation for 
small energy gaps in the FQHE regime. It is 
most likely attributable to residual disorder 
in the specimen, which leads to a smearing 
out of the energy gap. For FQHE states around 
$\nu=1/2$ an approximately $B$-independent gap 
reduction has been deduced from activation 
energy measurements on the primary $\nu=p/(2p \pm 1)$ 
FQHE sequence and was rationalized as a broadening 
of the associated CF Landau level \cite{du:prl94}. Even for 
such very high-mobility samples the value of 
the gap reduction is $\sim 2$~K. This value also 
agrees with scattering rates determined at $\nu=1/2$. 
We must assume that a similar gap reduction 
mechanism is at work in the second Landau level 
and that its magnitude is not unlike the 
magnitude around $\nu=1/2$. Accepting such a rationale, 
we are lead to conclude that, in fact, 
$\Delta_{5/2} \approx \Delta_{7/3} \approx \Delta_{8/3}$ 
and all three gaps are on the order of $\sim$ 2~K. 
This brings the value of experiment in the 
range of theory, but the result is at 
variance with the theoretically derived 
ratios between the gaps \cite{morf:prl98}. The above 
line of thought has a more general implication. 
Accepting in general a gap reduction 
of $\sim$ 2~K for FQHE gaps due to remnant
disorder even in the highest mobility samples, 
one would have to conclude that presently 
measurements of activation energies much 
less than $\sim$ 2~K only establish the existence 
of true gaps of $\sim$ 2~K. Comparison between 
measured activation energies of different 
fraction becomes less meaningful, although 
it should remain an acceptable indicator for 
whether a specific gap increases or 
decreases in response to a perturbation. 

In summary, we have conducted ultra-low temperature 
experiments on the FQHE states in the second 
Landau level, at $\nu=5/2$ and its vicinity. 
For the first time, the longitudinal resistance 
assumes vanishingly low values at this ominous 
even-denominator filling factor and the 
concomitant Hall plateau is quantized to 
$2h/5e^2$ to an accuracy better than $2 \times 10^{-6}$. 
Neighboring minima at $\nu=7/3$ and $\nu=8/3$ also 
show vanishing resistance and Hall quantization 
to their respective quantum number. Activation 
energy measurement on all three fractions 
yield $\Delta_{5/2}=0.11$~K, $\Delta_{7/3}=0.10$~K, 
and $\Delta_{8/3}=0.055$~K, 
yet we deduce, that the underlying, true energy 
gaps for all three FQHE states are very 
similar and of magnitude $\Delta \sim 2$~K.

We would like to thank N. Bonesteel and R. R. Du 
for useful discussions. A portion of this work 
was performed at the NHMFL which is supported by 
NSF Cooperative Agreement No. DMR-9527035 and 
by the state of Florida. D.C.T. and W.P. 
are supported by NSF and the DOE.

\begin{figure}
\vspace{0.2cm}
\centerline{\psfig{figure=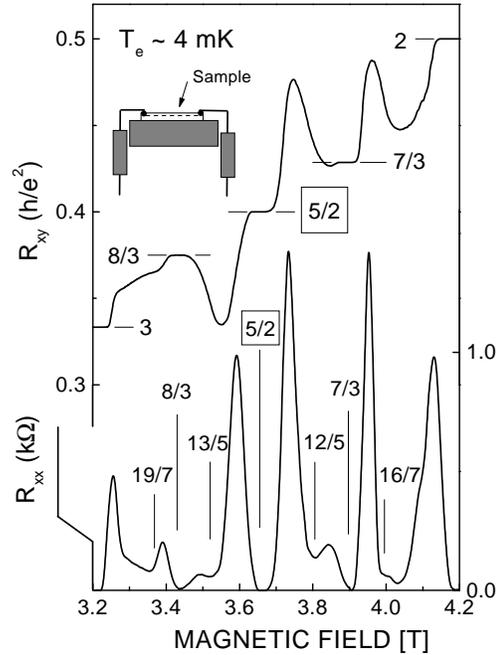,width=7.5cm,angle=0}}
\vspace{0.2cm}
\caption{
Hall resistance Rxy and longitudinal 
resistance Rxx at an electron temperature 
$T_e \approx 4.0$~mK. Vertical lines mark the Landau level 
filling factors. The inset shows a schematic 
of the sample with attached sintered silver 
heat exchangers (gray) to cool the 2DES.}
\vspace{0.2cm}
\end{figure}

\begin{figure}
\vspace{8.2cm}
\centerline{\psfig{figure=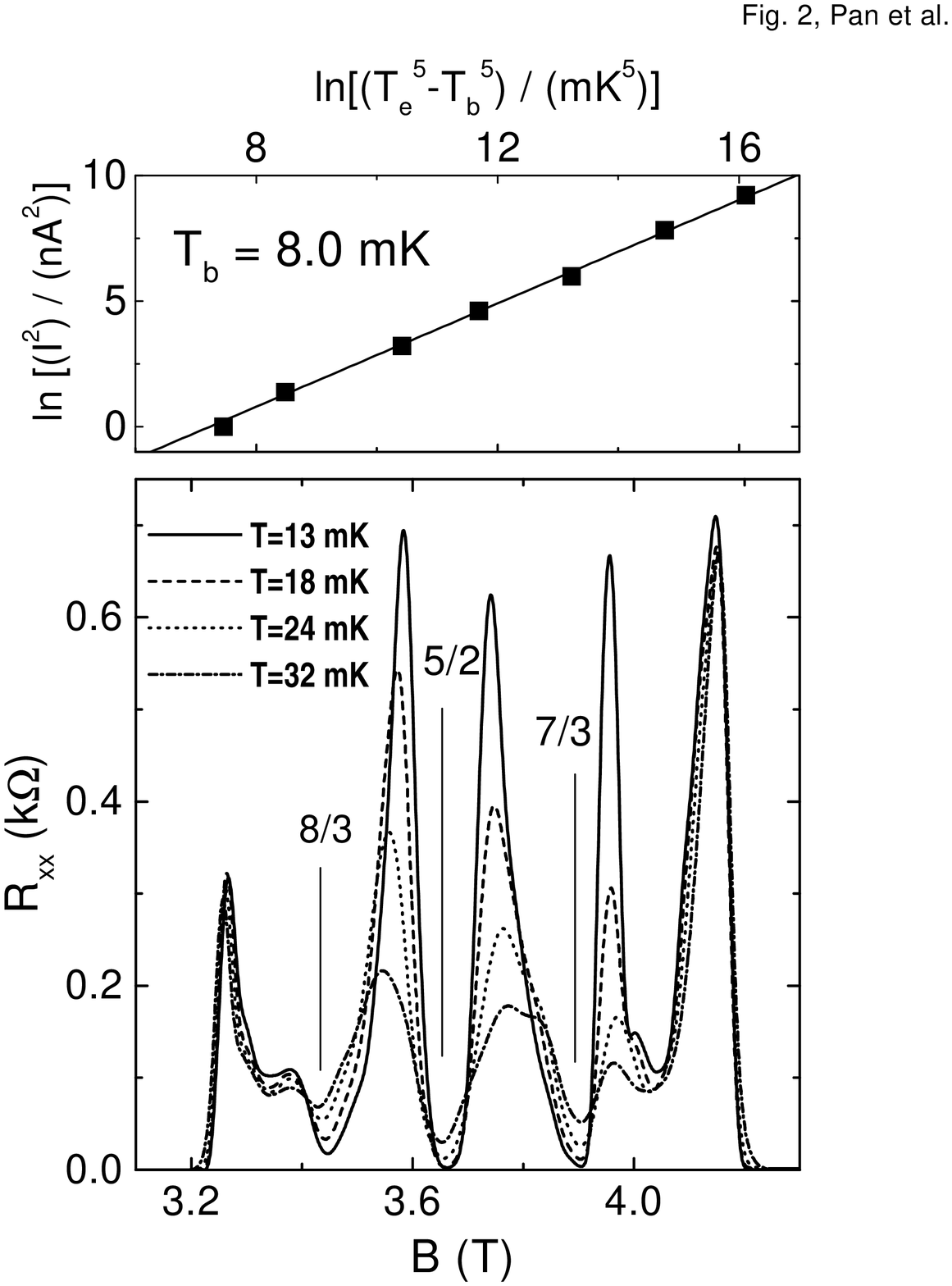,width=7.5cm,angle=0}}
\vspace{0.2cm}
\caption{
Lower panel: the temperature evolution of 
$R_{xx}$ between $\nu = 2$ and $\nu = 3$. 
Upper Panel: ln($I^2$) vs. ln($T_{e}^5-T_{b}^5$). 
$I$ is current in nA. $T_e$ is electron temperature 
and $T_b$ is bath temperature, in mK.} 
\vspace{0.2cm} 
\end{figure}

\begin{figure}
\vspace{8.2cm}
\centerline{\psfig{figure=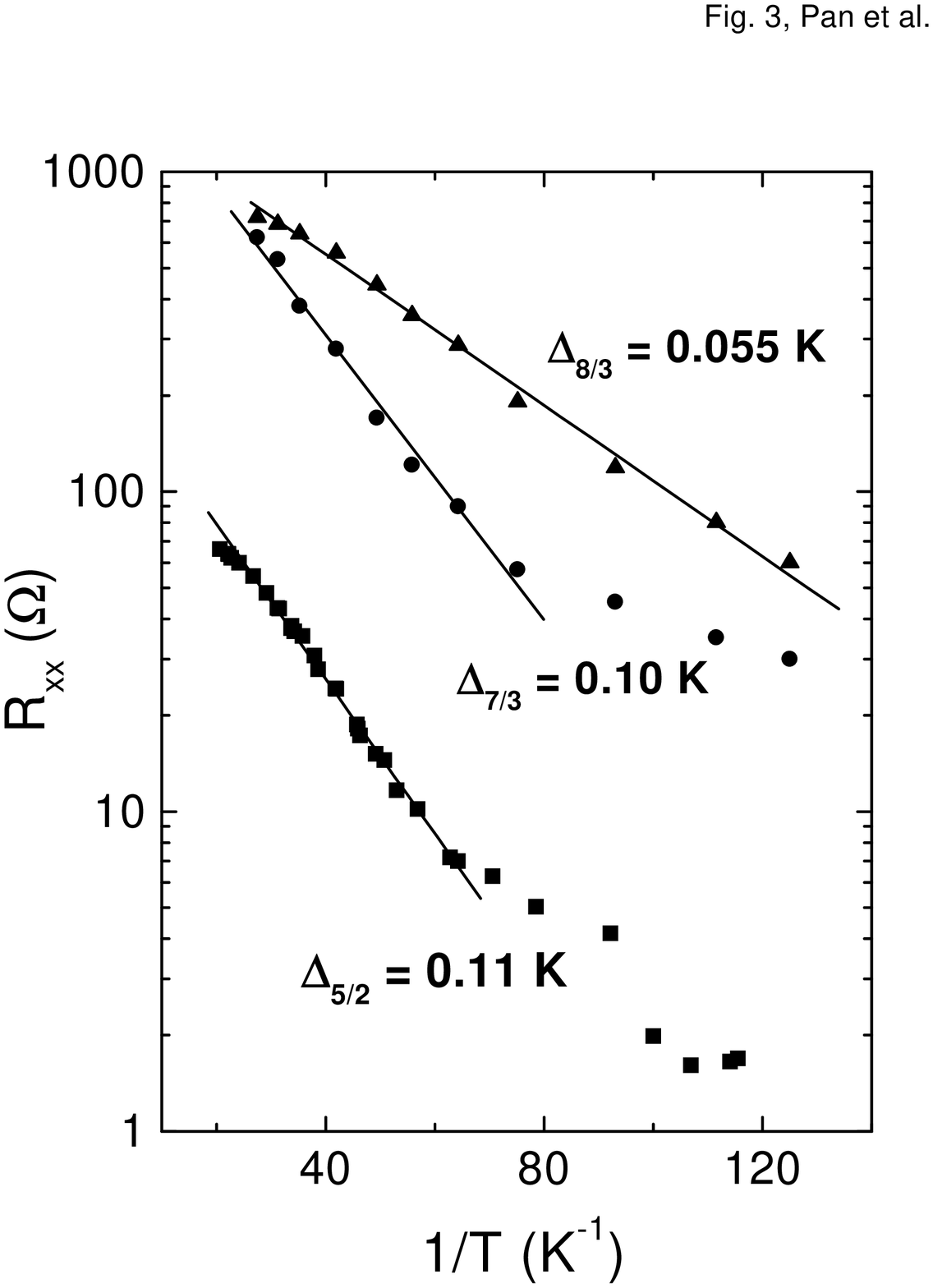,width=7.5cm,angle=0}}
\vspace{0.2cm}
\caption{
Activation energy plots for 
$R_{xx}$ at the $\nu$ = 5/2, 7/3, and 8/3 minima 
for 8.0~mK $< T_b <$ 50.0~mK. The data for $\nu=7/3$ 
and $\nu=8/3$ are multiplied by 10 for clarity. 
Activation energy gaps, $\Delta_{\nu}$, are determined 
from the slope of the extended linear regions.}
\vspace{0.2cm}
\end{figure}

\end{document}